\documentclass[showpacs,preprintnumbers,amsmath,amssymb,twocolumn,aps,prl]{revtex4}

\usepackage{graphicx}

\begin{document}

\title{Non-universal results induced by diversity distribution in coupled excitable systems}
\author{Luis F. Lafuerza, Pere Colet, Raul Toral}

\affiliation{IFISC, Instituto de F{\'\i}sica Interdisciplinar y Sistemas Complejos, CSIC-UIB,  Campus UIB, E-07122 Palma de Mallorca, Spain  }

\date{\today}


\begin{abstract}
We consider a system of globally coupled active rotators near the excitable regime. 
The system displays a transition to a state of collective firing induced by disorder. We show that this transition is found generically for any diversity distribution with well defined moments. Singularly, for the Lorentzian distribution (widely used in Kuramoto-like systems) the transition is not present. This warns about the use of Lorentzian distributions to understand the generic properties of coupled oscillators.
\end{abstract}

\maketitle
Synchronization phenomena play a prominent role in many branches of science \cite{pikowsky}. They have been analyzed in terms of phase models  which successfully describe systems of weakly coupled limit cycle oscillators. In particular, the Kuramoto model \cite{kuramoto1} has become a paradigm for the study of synchronization (for reviews see \cite{pikowsky,acebron,strogatzreview}). It shows how synchronized behavior can appear when the competitive effects of coupling and diversity among the individual units are present.

Amongst other possible sources, diversity in the oscillators is usually introduced by taking their natural frequencies from a probability distribution. Although, on general grounds (central limit theorem), this distribution should be well approximated by a Gaussian form, theoretical studies usually consider a Lorentzian form since it allows for an easier analytical treatment. It is generally believed that the main results concerning the global synchronization properties are qualitatively independent of the precise form of the distribution as long as it is symmetric and unimodal. In this Letter, however, we show that a variant of the Kuramoto model displays or not a reentrant diversity-induced transition into a state of collective firing, depending on the type of distribution used. This transition is present in all the (symmetric and unimodal) distributions studied, but not in the case of the Lorentzian. The non-generic behavior of the system with a Lorentzian distribution of natural frequencies warns about the indiscriminate use of some recently proposed methods \cite{ott} in order to understand generic properties of coupled oscillators.

We consider the following variant of the Kuramoto model which describes the dynamics of an ensemble of globally coupled active rotators $\phi_{j}(t), j=1,...,N$ \cite{kuramoto2}:
\begin{equation}
 \dot{\phi}_{j}=\omega_{j}-\sin\phi_{j}+\frac{K}{N}\sum_{l=1}^{N}\sin(\phi_{l}-\phi_{j}).
 \label{system}
\end{equation}
A natural frequency $\omega_{j}<1$ (respectively, $\omega_{j}>1$) corresponds to an excitable (respectively, oscillatory) behavior of the rotator $j$ when it is uncoupled. $K$ is the coupling intensity.  Diversity is introduced by considering that the $\omega_{j}$'s are distributed according to a probability density function $g(\omega)$, with mean value $\overline{\omega}$ and variance $\sigma^{2}$. The model is equivalent to the Kuramoto model with zero average frequency and an external periodic driving of frequency $- \overline{\omega}$, as it can be easily seen with the change of variables $\phi_{j}\rightarrow\phi_{j}-\overline{\omega}t$.
Throughout the paper, besides the well-known Gaussian and uniform distributions, we will be considering a general family of distributions $L_n^m(\omega)$, for $n> 0,mn> 1$, defined as:
\begin{equation}
L_n^m(\omega)=\frac{n\Gamma(m)}{2\Gamma(m-1/n)\Gamma(1/n)}\cdot\frac{\Delta^{nm-1}}{(|\omega-\overline{\omega}|^{n}+\Delta^{n})^{m}}.
\end{equation}
The variance is finite only for \makebox{$mn>3$} and it is given by 
\makebox{$\sigma^2=\Delta^2\frac{\Gamma(m-3/n)\Gamma(3/n)}{\Gamma(m-1/n)\Gamma(1/n)}$}. 
The Lorentzian distribution corresponds to $L_2^1(\omega)$ and has, hence, an infinite variance, although we still will use $\Delta$ as a measure of diversity.

To characterize the collective behavior we use the time-dependent complex variable $r(t)$ \cite{kuramoto1,kuramoto2}:
\begin{equation}
 r(t)=\frac{1}{N}\sum_{j=1}^{N}e^{i\phi_{j}(t)}.
\end{equation}
The Kuramoto order parameter $\rho\equiv\langle |r(t)|\rangle$, where $\langle \cdots \rangle$ denotes time average, is known to be a good measure of collective synchronization in coupled oscillators systems, i.e. $\rho\simeq1$ when the oscillators synchronize ($\phi_{j}\simeq\phi_{l}, \forall j,l$), and $\rho\simeq0$ for desynchronized behavior.

For $\overline \omega\lesssim1$ the  system displays three different regimes: (i) for small diversity, almost all units are at rest at similar fixed points ($r(t)$ does not depend on time and $\rho\approx 1$); (ii) increasing diversity one enters a dynamical state in which a macroscopic fraction of units fire at (roughly) the same time ($r(t)$ is time dependent while $\rho$ is still close to 1); (iii) for even larger diversity, the system reenters a desynchronized state ($\rho$ small and $r(t)$ time independent). To discriminate between static entrainment and collective firing, regimes (i) and (ii), we use the order parameter introduced by Shinomoto and Kuramoto \cite{shinomoto}:
\begin{equation}
 \zeta=\langle |r(t)-\langle r(t)\rangle|\rangle,
\end{equation}
which differs from zero only for collective firing.

\begin{figure}[h]
\centering
\includegraphics[scale=0.25,angle=0,clip]{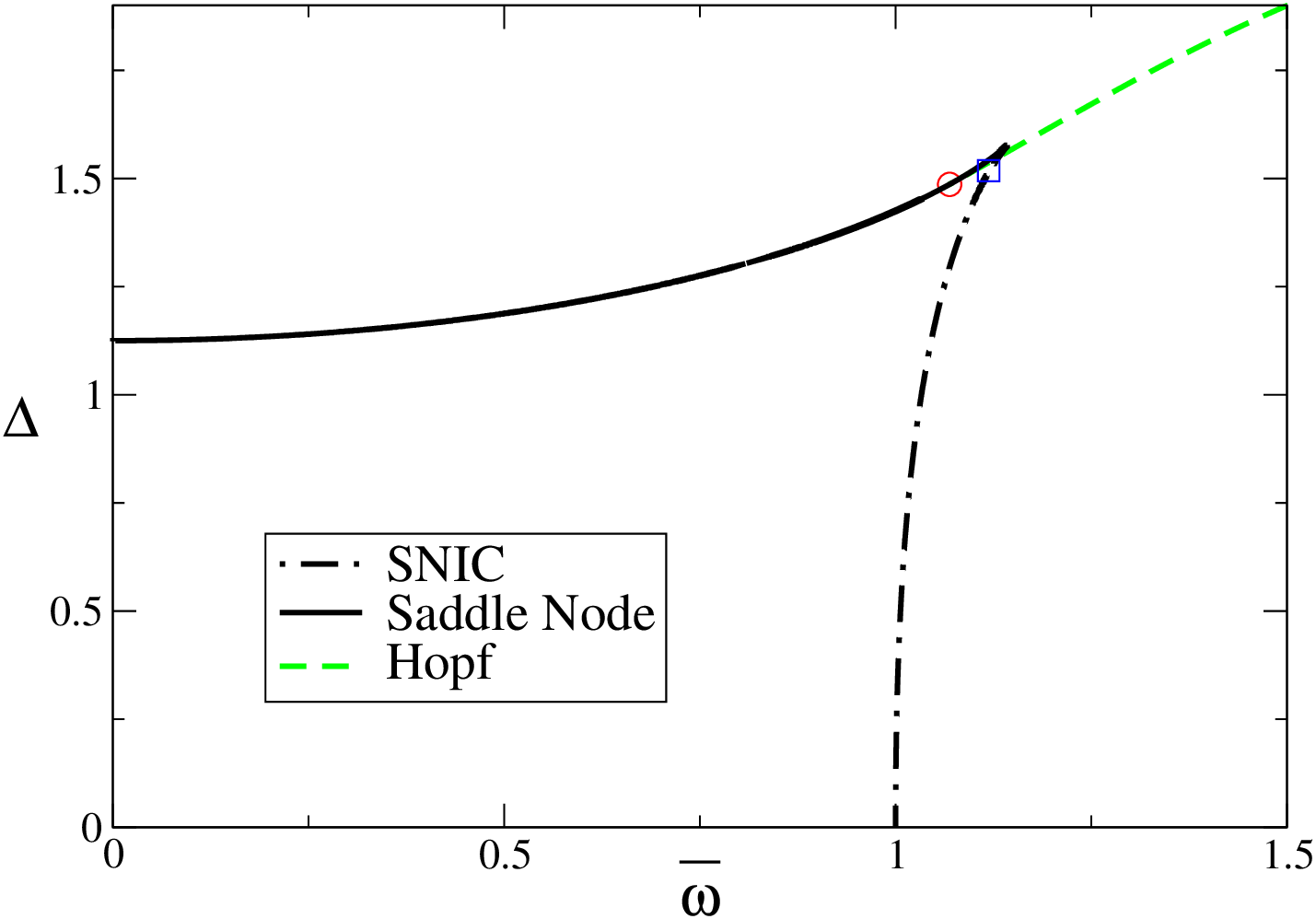}
\caption{Bifurcation diagram of the  model defined by Eqs. (\ref{system}) for a Lorentzian distribution of frequencies and $K=5$. 
There is also a saddle-loop bifurcation line, not shown, that goes from the Takens-Bogdanov point (circle) to the saddle-node separatrix loop point (square) where SNIC line starts \cite{strogatz}. Collective firing takes place in the region below the Hopf and right of the SNIC.
\label{bifurcation}} 
\end{figure}

An approximate theory to describe these three regimes was developed in \cite{tessone}. The theory was independent of the form of the natural frequencies distribution and was also applicable to identical units subject to noise. A recent method developed by Ott and Antonsen \cite{ott,ott2} allows to solve exactly this model (and a large family of related ones) in the infinite number of oscillators limit and in a number of cases that include the Lorentzian $g(\omega)$. Childs and Strogatz \cite{strogatz} used this method to obtain the full bifurcation diagram of the complex variable $r(t)$ for the Lorentzian distribution. Contrarily to the results of \cite{tessone}, their exact solution implies that there is no transition to collective firing increasing the diversity for $\overline \omega<1$. The non-existence of the transition can be derived from the bifurcation diagram in the $\overline{\omega}-\Delta$ space obtained using the ideas of \cite{strogatz}, see Fig.\ref{bifurcation}. Regime (ii) takes place for the parameter region located to the right of the SNIC (saddle node on the invariant circle) bifurcation line and below the Hopf line. For $\overline \omega <1$ increasing $\Delta$ one never enters in this region, so there is no 
diversity-induced transition to collective firing. This situation is generic for all values of $K$, since it can be shown that the SNIC bifurcation always starts at $\overline \omega=1,\Delta=0$ with positive slope. The model was also studied for the Lorentzian case with a different approach in \cite{antonsen} and the same results were found.

We will give now the main sketches of the Ott-Antonsen method. Quite generally, we will show that, in fact, the method can be successfully applied to any non-singular distribution $g(\omega)$. Let $f(\omega,\phi,t)$ be the density of oscillators with frequency $\omega$ and phase $\phi$ at time $t$. This function obeys the continuity equation (conservation of the number of oscillators):
\begin{equation}
 \frac{\partial f(\omega,\phi,t)}{\partial t}+\frac{\partial}{\partial \phi}\left[\dot{\phi}(\omega,\phi,r)f\right]=0,
\end{equation}
with $\dot{\phi}(\omega,\phi,r)=\omega-\sin\phi+\Re(re^{-i\phi})$.
If the coefficients of the Fourier expansion:
\begin{equation}
f(\omega,\phi,t)=\frac{g(\omega)}{2\pi}\left[ 1+\sum_{m=1}^{\infty}\left[f_{m}(\omega,t)e^{im\phi}+c.c.\right]\right]
\end{equation}
($c.c.$ denotes complex conjugate), satisfy the ansatz
\begin{equation}
f_{n}(\omega,t)=\alpha(\omega,t)^{n}\label{oaansatz},
\end{equation}
then $\alpha(\omega,t)$ and $r(t)=\int d\omega \int d\phi \,e^{i\phi}f(\omega,\phi,t)$ obey the integro-differential equations:
\begin{eqnarray}
&& \frac{\partial \alpha}{\partial t}+i\omega\alpha+\frac{1}{2}\{\left[Kr+1\right]\alpha^{2}-Kr^{*}-1\}=0,
 \label{otteq}
\\ &&r(t)=\int d\omega\,\alpha(\omega,t)^{*}g(\omega).
 \label{req}
\end{eqnarray}
The manifold defined by (\ref{oaansatz}) is invariant under the system evolution, so if that condition is fulfilled by the initial condition, it is fulfilled afterwards. Moreover, in \cite{ott2} it is shown that for a Lorentzian $g(\omega)$ the long time evolution of $r(t)$ is always described by this reduced manifold. 

If $g(\omega)$ has a finite set of poles $\hat \omega_1,\hat \omega_2,\dots$  outside the real axis (as is the case for $L_n^m(\omega)$ for even $n$ and integer $m$, including the Lorentzian $L_2^1(\omega)$, and $\alpha(\omega,t)$ satisfies certain analyticity conditions, one can obtain (\ref{req}) by contour integration. Then $r(t)$ can be written is terms of $\alpha_k(t)\equiv\alpha(\hat \omega_k,t)$ and one can obtain a closed set of ordinary differential equations for $\alpha_k(t)$. In the case of poles with multiplicity larger than one, $r(t)$ depends also on the partial derivatives with respect to $\omega$, $\alpha_k^s(t)\equiv \alpha^{(s)}(\hat \omega_k,t)$. Equations for these new functions $\alpha_k^s(t)$ can be obtained by differentiating Eq.(\ref{otteq}) with respect to $\omega$. For an arbitrary distribution $g(\omega)$, we can obtain an approximate evolution of the system by evaluating integral (\ref{req}) using a finite, though large, set of values of  $\omega$ and integrating numerically (\ref{otteq}) for each one of these frequencies.

\begin{figure}[h]
\centering
\includegraphics[scale=0.38,angle=0,clip]{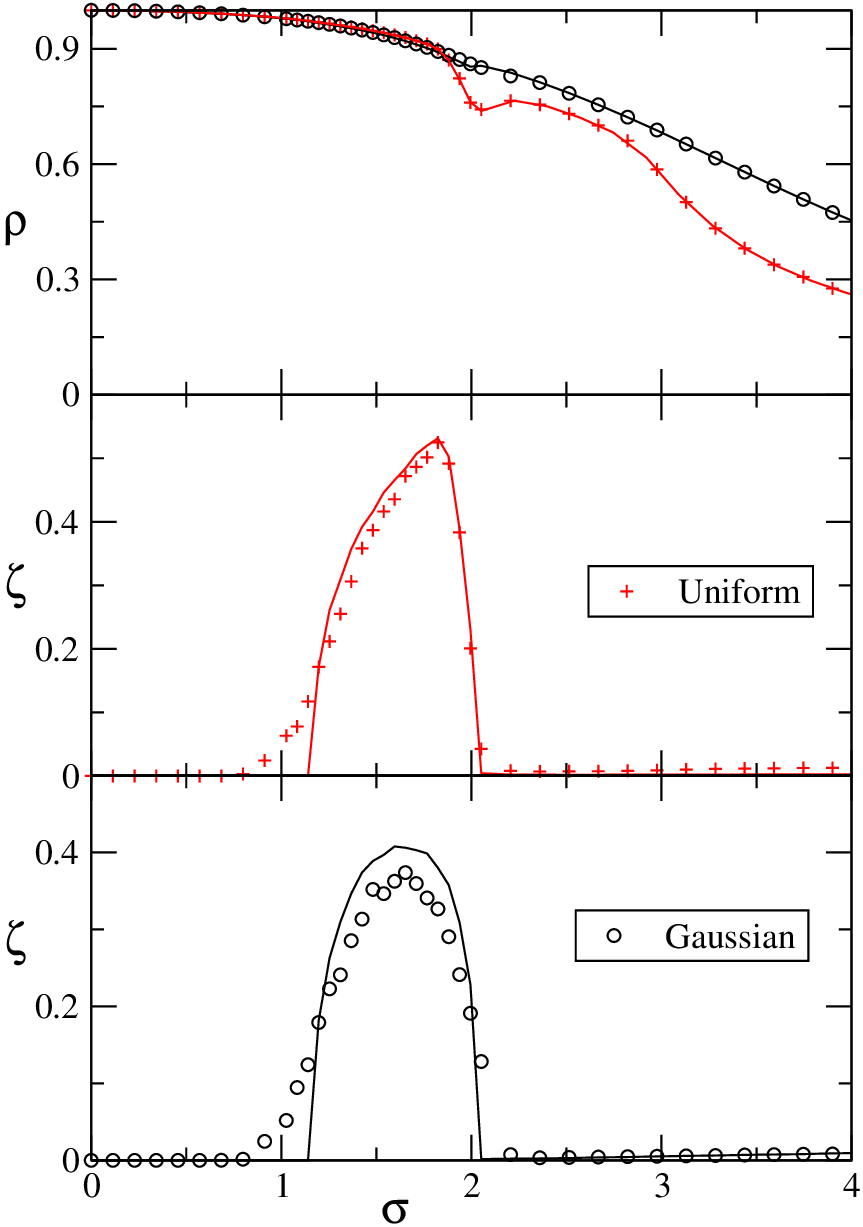}
\caption{Order parameters $\rho$ and $\zeta$ as a function of the diversity $\sigma$ coming from numerical simulations of the full system of Eqs.(\ref{system}) for $N=10^4$ units in the cases of a Gaussian (dots) and uniform (crosses) distribution of frequencies. In each case, the solid lines are the result of the application of the Ott-Antonsen method using $10^4$ values of $\omega$ for the numerical integration of Eq.(\ref{req}). $K=5$ and $\overline{\omega}=0.97$
\label{gaussuniform}}
\end{figure}
\begin{figure}[h]
\centering
\includegraphics[scale=0.38,angle=0,clip]{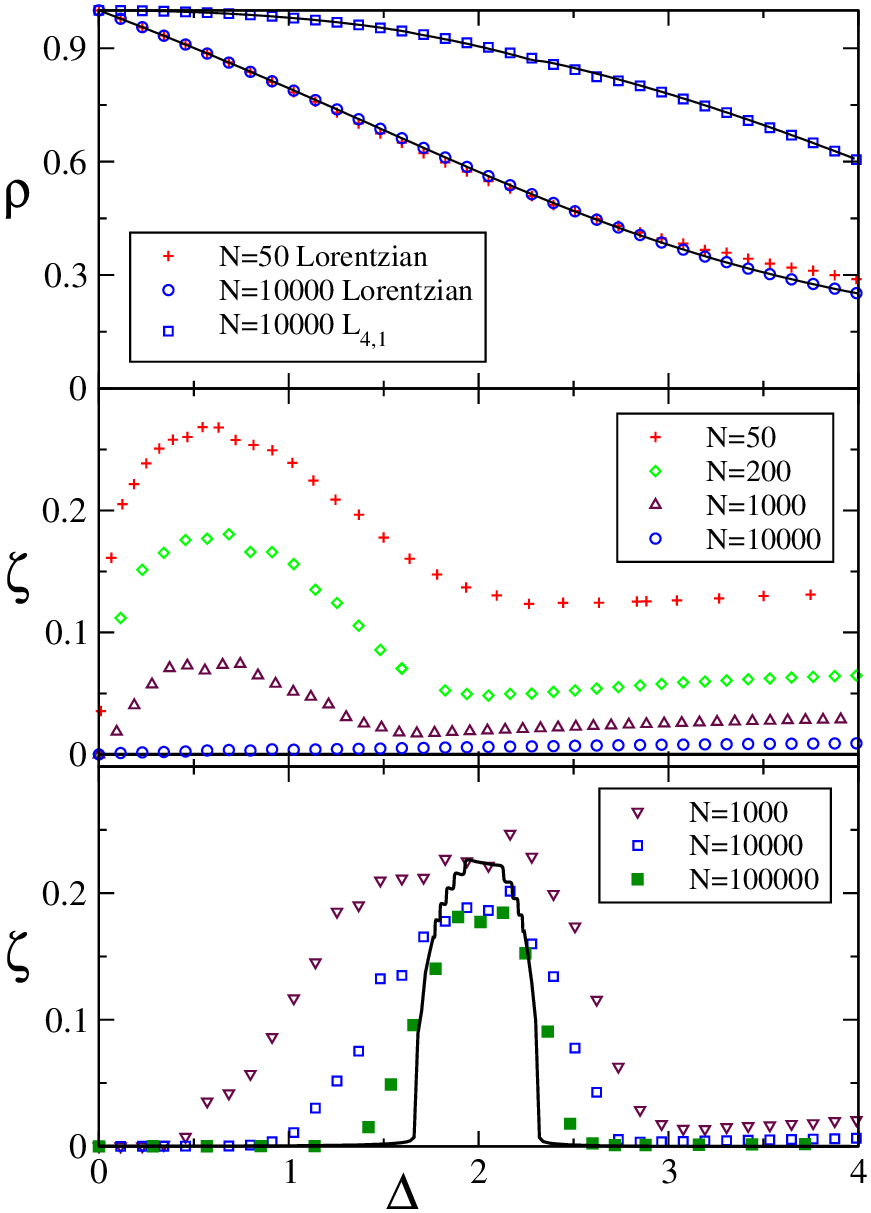}
\caption{Same as Fig.\ref{gaussuniform} for Lorentzian $L_{2}^{1}(\omega)$ (upper and middle panels) and $L_{4}^{1}(\omega)$ (upper and lower ones) distributions. Note that for the Lorentzian distribution $L_{2}^{1}(\omega)$  the Ott-Antonsen method predicts that $\zeta$ is zero for all values of the diversity $\Delta$ while the numerical simulations for finite system size $N$ show a rounded-off transition. \label{lorentzians}} 
\end{figure}

In Figs. \ref{gaussuniform} and \ref{lorentzians}  we show the order parameters as a function of the diversity for several frequency distributions, obtained by direct simulation of Eqs. (\ref{system}) and by applying Ott-Antonsen's method as described above. For the Gaussian and uniform distributions, but not for the Lorentzian, the regimen of collective firing (signaled by a nonzero value of the parameter $\zeta$) is present for intermediate values of the diversity. Beyond the cases shown in the figures, the transition is also present for other symmetric distributions such as the exponential $g(\omega)=\frac{1}{\sigma\sqrt{2}}e^{-\frac{\sqrt{2}}{\sigma}|\omega-\overline{\omega}|}$, or the family $L_{n}^{m}(\omega)$ for $mn \geq 3$. Even $L_{3}^{1}(\omega)$ which has infinite variance (but well-defined first moment) presents the reentrant diversity-induced transition (for values of $\overline\omega$ close enough to one). Also, if we truncate the Lorentzian distribution at some finite value of $\omega$, i.e. set $g(\omega)=0$, if $|\omega-\overline{\omega}|>C$, the system shows this reentrant transition (we checked for $C$=50$\Delta$).  Furthermore, Fig. \ref{lorentzians} shows that for finite-size systems with a Lorentzian distribution a maximum in $\zeta$ is indeed present, being quite visible up to a few thousand of units.  In fact, Lorentzian distributions in systems with a finite number of units are effectively truncated, truncation that disappears in the limit $N\rightarrow \infty$. We conclude that the existence of the transition to common firing is a truly generic phenomenon and the results obtained using a Lorentzian distribution in the infinite system size limit are rather pathological and misleading.

\begin{figure}[h]
\centering
\includegraphics[scale=0.4,angle=0,clip]{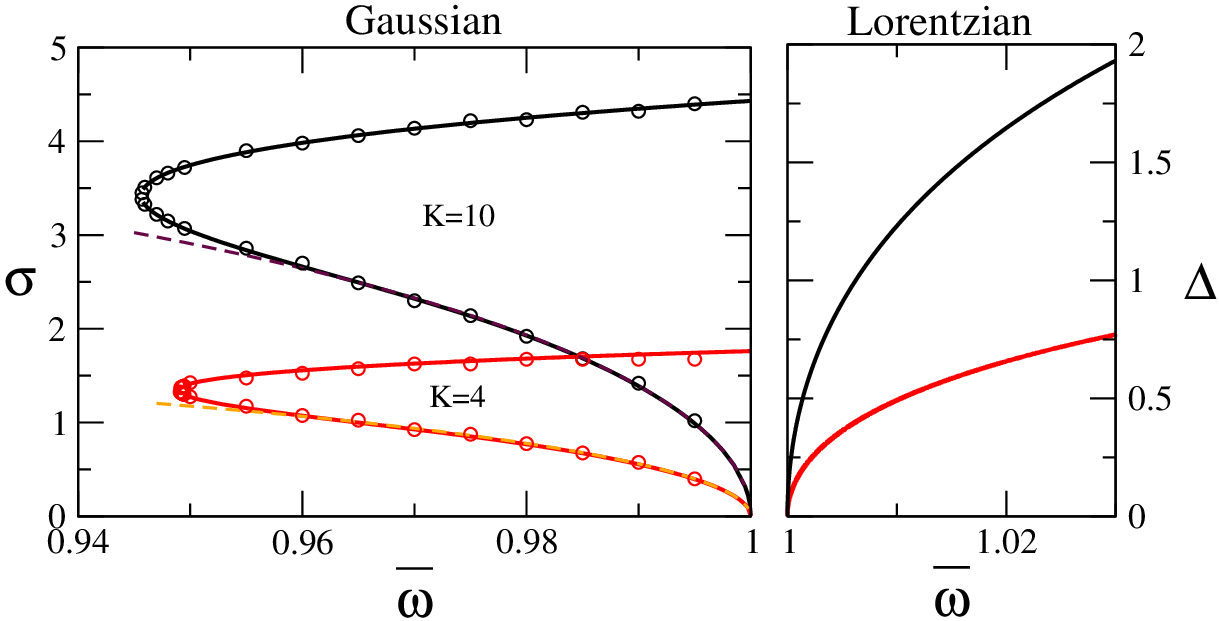}
\caption{(Partial) Bifurcation diagram for Gaussian (left) and Lorentzian (right) distributions. For the  Gaussian, circles correspond to the Ott-Antonsen method solved numerically as in Fig. (\ref{gaussuniform}), solid lines to the numerical solution of Eqs.(\ref{consistent1}-\ref{consistentb}) and dashed lines to Eq.(\ref{approxg}). For the Lorentzian distribution the lines correspond to the Ott-Antonsen method. \label{bifurcationgauss}}
\end{figure}

We introduce now an alternative approach to determine the parameters for which the system shows collective firing. Notice that $r(t)$ is time independent in regimes (i) (units at rest) and (iii) (desynchronized state) but not in (ii). Following previous analysis in the limit of infinite units \cite{sakaguchi}, we write $r(t)=\rho e^{i\Psi}$ and, assuming that $r(t)$ is time independent, derive the following equations for the global amplitude $\rho$ and phase $\Psi$:
\begin{eqnarray}
 \rho\sin\Psi=\overline{\omega}-\int_{|\omega|>b}\omega\sqrt{1-\frac{b^{2}}{\omega^{2}}}g(\omega)d\omega\equiv f_1(b),\label{consistent1}\\
K\rho^{2}+\rho\cos\Psi=\int_{-b}^{b}\sqrt{b^{2}-\omega^{2}}g(\omega)d\omega  \equiv f_2(b),\label{consistent2}
\end{eqnarray}
with $b=\sqrt{1+K^{2}\rho^{2}+2K\rho\cos\Psi}$. It is possible to obtain a closed self-consistent relation for $b$:
\begin{equation}
b=\frac{K(f_{1}^{2}+f_{2}^{2})}{bf_2\pm\sqrt{f_2^2+(1-b^2)f_1^2}}\equiv h_{\pm}(b).\label{consistentb}
\end{equation}
For parameters corresponding to regime (ii) Eq. (\ref{consistentb}) has only one solution corresponding to
an unstable fix point. When the branches $h_+(b)$ and $h_-(b)$ meet two new solutions arise. This signals a SNIC bifurcation where a stable fix point and a saddle are created. 
Imposing $h_-(b)=h_+(b)=b$, one can determine the line $\sigma(\overline{\omega})$ at which the transition takes place. Noticeably, on the SNIC  one obtains $\Psi=\pi/2$ independent of the form of the distribution $g(\omega)$ or the values of the parameters. As shown in Fig. \ref{bifurcationgauss} the results of this approach coincide with those of the Ott-Antonsen method.

For the Gaussian distribution,  the SNIC line that limits the collective firing regime starts at $\overline{\omega}=1, \sigma=0$, with a negative slope. Therefore for $\overline{\omega}<1$, as $\sigma$ increases one finds first a stable steady state for $r$ (regime (i)), then crosses the SNIC lower boundary entering in regime (ii) and finally crosses the reentrant upper boundary entering in regime (iii) where a stable steady state is present again. Region (ii) moves upwards and broadens increasing the coupling $K$.  The overall phase diagram still has a Takens-Bogdanov point as in Fig. 1. Decreasing $K$ this point moves down and to the left. For low values of $K$, for example $K=1$, the upper boundary of the collective firing region is limited by the Hopf rather than by the reentrant SNIC, nevertheless the diversity-induced transition is still present. This reentrance behavior is generic for all distributions with well defined moments.

The Lorentzian distribution behaves differently. The SNIC starts at $\overline{\omega}=1,\Delta=0$ with positive slope. Although apparently this is a small quantitative difference and the phase diagrams may be topologically equivalent, there are important qualitative consequences and, in particular, no diversity-induced transition to collective firing exists\footnote{For $\overline{\omega}\gtrsim1$ there is a small region limited by the homoclinic and the Hopf bifurcations where there is bistability between a steady state and one with collective motion. However in the collective motion state, $r$ moves slightly around a fixed value rather than performing collective firing, and the basin of attraction of this state is very small, so this is indeed different from the regime we are considering.}

For distributions which decay fast enough we can obtain an analytical approximation for the line $\sigma(\bar{\omega})$ at which the transition to collective firing appears. We assume that $g(\omega)\simeq0$ for  $|\omega|>b$. As $b\ge |K\rho-1|$, this approximation turns out to be appropriate for large values of $K$. Using $\Psi=\pi/2$ one gets $\rho=\overline{\omega}$ from Eq. (\ref{consistent1}).  Inserting this in (\ref{consistent2}) and expanding the integrand to second order in $\omega/b$ we obtain
\begin{equation}
 \sigma(\bar{\omega})=\sqrt{\overline{\omega}^{2}(2K^{2}-2K\sqrt{1+K^{2}\overline{\omega}^{2}}-1)+2}.\label{approx} 
\end{equation}
From Eq. (\ref{approx}) one can obtain $d \sigma/d\bar{\omega}$. At the origin of the SNIC line ($\bar{\omega}=1$), the derivative takes a negative value for any $K>0$ and tends to $-\infty$ as $-K^3$ for $K\rightarrow \infty$. Notice that this is independent of the form of $g(\omega)$ provided the tails decay fast enough. Higher order corrections do depend on the specific form $g(\omega)$. For the Gaussian the next order gives:
\begin{eqnarray}
&&\sigma(\bar{\omega})^{2}=\frac{-\overline{\omega}^{2}(3+2K^{2})-2}{3}+\label{approxg}\\ 
&&\frac{\sqrt{\overline{\omega}^{2}[6\overline{\omega}^{2}-24K(1+K^{2}\overline{\omega}^{2})^\frac{3}{2}]+28(1+K^{2}\overline{\omega}^{2})^{2}}}{3}.\nonumber
\end{eqnarray}
This approximation, plotted as dashed lines in Fig. \ref{bifurcationgauss}, describes well the location of the transition to firing between regimes (i) and (ii) for the case of a Gaussian distribution.

In summary, globally coupled active rotators display a diversity induced transition to a state of collective firing. We have shown that this is a genuine transition that is found for any disorder distribution with well defined moments. Curiously enough, the transition is not present for a Lorentzian distribution, for which the first moment integral is only defined as a principal value. 
We have also found that these non-generic results given by the Lorentzian distribution also occur in another well-known excitable system, an ensemble of coupled FitzHugh-Nagumo \cite{nagumo} units, for which the reentrant diversity-induced transition is present as well for distributions such as Gaussian or uniform, but not for the Lorentzian one.

While the Lorentzian distribution seems to be suitable to study generic properties of the original Kuramoto model, it is not appropriate for excitable systems and beyond those may not be adequate for other systems. This is a clear warning about its use in analytical approximations intended to draw conclusions on the generic properties of coupled oscillators.  

{\textbf{Acknowledgments:}}
We acknowledge financial support by the MICINN (Spain) and FEDER (EU) through project FIS2007-60327. L.F.L. is supported by the JAEPredoc program of CSIC.

\end{document}